\documentclass[11pt,justify]{article}
\usepackage{times, wrapfig}
\usepackage{geometry,natbib,graphicx}
\usepackage[utf8]{inputenc}
\usepackage{enumitem,amssymb}
\usepackage{ragged2e}
\newlist{thematic}{itemize}{8}
\setlist[thematic]{label=$\square$}
\usepackage{pifont}

\begin{document}
\raggedright
\Large
{\bf Astro2020 Science White Paper \linebreak

The Next Generation Celestial Reference Frame }\linebreak
\normalsize

\noindent \textbf{Primary Thematic Area:} $\boxtimes$    Galaxy Evolution\\
%\vspace{4pt}
{\small \noindent \textbf{Secondary Thematic Areas:} $\boxtimes$ Planetary Systems \hspace*{10pt} $\boxtimes$ Star and Planet Formation \hspace*{10pt}
$\boxtimes$ Formation and Evolution of Compact Objects \hspace*{10pt} $\boxtimes$ Cosmology and Fundamental Physics 
 \hspace*{10pt}  $\boxtimes$  Stars and Stellar Evolution \hspace*{10pt} $\boxtimes$ Resolved Stellar Populations and their Environments \hspace*{10pt} 
    $\boxtimes$  Multi-Messenger Astronomy and Astrophysics}  \linebreak

 %\vspace{18pt}
 \textbf{Principal Author:}

Megan C.\ Johnson \hfill  {\small (202) 762-1488} \\
{\small United States Naval Observatory \hfill megan.johnson@navy.mil} \linebreak

%\vspace{8pt}
\textbf{Co-authors:} (names and institutions)\\
%\linebreak
{\small
Frank Schinzel$^1$, %, National Radio Astronomy Observatory\\
Jeremy Darling$^2$, %, University of Colorado\\
Nathan Secrest$^3$, %, United States Naval Observatory}
Bryan Dorland$^3$, %, United States Naval Observatory\\
Alan Fey$^3$, %, United States Naval Observatory\\
Leonid Petrov$^4$, %, NASA Goddard Space Flight Center\\
Anthony Beasley$^1$, %, National Radio Astronomy Observatory \\
Walter Brisken$^1$, %, National Radio Astronomy Observatory\\
John Gipson$^{4,5}$, %, NVI, Incorporated, NASA Goddard Space Flight Center \\
David Gordon$^{4,5}$, %, NVI, Incorporated, NASA Goddard Space Flight Center \\
Lucas Hunt$^3$, %, United States Naval Observatory\\
Joseph Lazio$^6$ } %, Jet Propulsion Laboratory\\

%\linebreak
\emph{\footnotesize $^1$National Radio Astronomy Observatory, $^2$University of Colorado,
$^3$United States Naval Observatory, $^{4}$NASA Goddard Space Flight Center,
$^{5}$NVI, Incorporated, $^6$Jet Propulsion Laboratory, California Institute of Technology}

 \begin{figure}[ht!]
   \centering
    \includegraphics[scale=0.2]{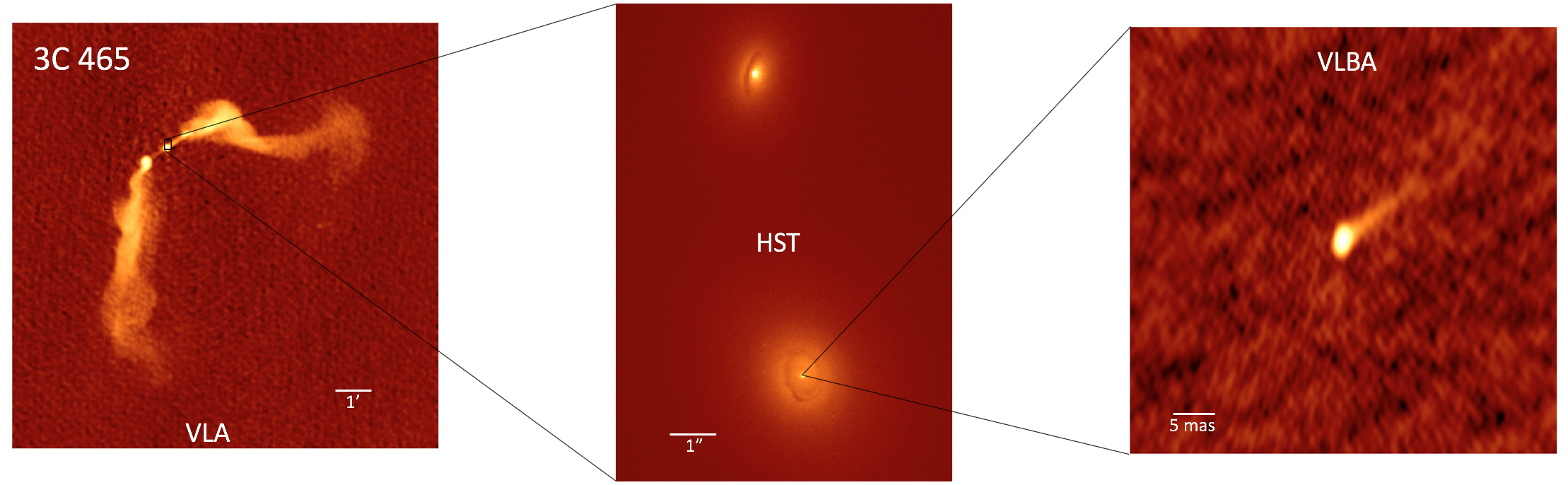}
\end{figure}

\vspace{2pt}
\textbf{Abstract:}
\vspace{-6pt}
\justify
Astrometry, the measurement of positions and motions of the stars, is one of the oldest disciplines in Astronomy, extending back at least as far as Hipparchus' discovery of the precession of Earth's axes in 190 BCE by comparing his catalog with those of his predecessors.  Astrometry is fundamental to Astronomy, and critical to many aspects of Astrophysics and Geodesy.  In order to understand our planet's and solar system's context within their surroundings, we must be able to to define, quantify, study, refine, and maintain an inertial frame of reference relative to which all positions and motions can be unambiguously and self-consistently described.  It is only by using this inertial reference frame that we are able to disentangle our observations of the motions of celestial objects from our own complex path around our star, and its path through the galaxy, and the local group. Every aspect of each area outlined in the call for scientific frontiers in astronomy in the era of the 2020-2030 timeframe will depend on the quality of the inertial reference frame.  In this white paper, we propose support for development of radio Very Long Baseline Interferometry (VLBI) capabilities, including the Next Generation Very Large Array (ngVLA), a radio astronomy observatory that will not only support development of a next generation reference frame of unprecedented accuracy, but that will also serve as a highly capable astronomical instrument in its own right.   Much like its predecessors, the Very Long Baseline Array (VLBA) and other VLBI telescopes, the proposed ngVLA will provide the foundation for the next three decades for the fundamental reference frame, benefitting astronomy, astrophysics, and geodesy alike.  

\pagebreak

\section{Introduction}

The International Astronomical Union (IAU) adopted the International Celestial Reference 
Frame (ICRF) as the Fundamental Celestial Reference Frame for all astronomy in January 1998.  
The ICRF is defined by radio positions of active galactic nuclei (AGNs) and uses the VLBI 
technique to maintain and improve its accuracy.  The ICRF-3 (4536 sources) is built on analysis of several decades of VLBI observations and
provides the best implementation of the CRF at sub-mas accuracies in radio wavelengths.
VLBA observations
accounted for at least 70\% of the sources and accuracy of the ICRF-3, and without a successor to the VLBA, monitoring
and maintenance of this number of ICRF-3 sources will not be possible.
%VLBA observations accounted for $\sim$70\% of the sources
%in the ICRF-3, and that without a successor to the VLBA,
%monitoring and maintenance of $\sim$70\% of the ICRF-3 will
%not be possible.

%\linebreak

In order to achieve an astrometric reference grid accurate to $\sim\rm \mu$as scales, we must 
be able to characterize, monitor, and improve AGN positions used in the CRF catalogs.  
The only demonstrated method by which modern astronomy has to achieve these goals continuously is through radio VLBI.  
%\linebreak

%The only method by which modern astronomy has to achieve these goals is through radio VLBI, and the Very Long Baseline Array (VLBA) currently, is essential to these activities.  Looking toward the future, it is highly desirable that the ICRF as realized by radio VLBI continues as the state-of-the-art celestial reference frame. 
%Thus, studying, understanding, and identifying the physical processes that contribute to the radio emission observed in the quasars that composite the ICRF is imperative to our ability to improve the reference frame.  

{\bf 
Development of the next generation of ICRF, along with the enabled science, will only be possible if the long baseline capabilities of the current VLBI systems are replaced with dramatically improved sensitivity and coverage capabilities of a next generation of long baseline radio telescopes.}

\section{Celestial Reference Frame as the Foundation of Astronomy}

\subsection{Radio and Optical Reference Frames}

  ESA's Gaia mission, the highly successful optical astrometry satellite, has already produced 
what is, effectively, an optical reference frame. It has measured over a billion stars with 
astrometric accuracies comparable to the ICRF for the brightest hundreds of millions 
of stars including over 9,000 AGNs directly observed with VLBI.
%\linebreak

  Initial comparisons have shown that for a sizable fraction of these objects, 7--15\%,
there are statistically significant positional offsets \citep[see, e.g.,][]{zac14, mak17, pet17a, fro18}.
%(see, for example, Zacharias, \& Zacharias(2014), 
%Makarov et al.(2017), Petrov, \& Kovalev(2017a), and Frouard et al.(2018)). 
These offsets 
appear to be due to a wide variety of possible astrophysical effects, such as background galaxy 
contamination or dual sources (see Figure on cover page), but the most common reason is the presence of optical jets at 1--100~mas scales
\citep{kov17, pla19a}
%(Kovalev et al.(2017), Plavin et al.(2019a)) 
that are too small to be detected even with the 
HST, but too large to affect position of the optical centroid. As a result, further improvement
of Gaia and VLBI accuracy will not reconcile positions of common sources, but will measure 
their positional offsets with greater accuracy \citep{pet17b} because the response 
of a power detector and an interferometer to an extended structure is fundamentally different.
As a consequence, the accurate Gaia optical position cannot be transferred to the radio range
and vice versa, if accuracy better than 1--3~mas is required. Furthermore, it has been 
recently established \citep{pet19} that these offsets are variable and result in 
statistically significant Gaia AGN proper motions induced by the variability of mas-scaled 
optical structures.
%\linebreak

  Although, on average, the radio and optical reference frames are aligned at a level of ten
$\mu$as; positions of individual objects diverge at a much higher level. Therefore, radio
and optical CRFs have to co-exist: applications of VLBI techniques, such as differential
radio astrometry, space navigation, and geodesy, have to rely on radio reference frames.
Applications of the optical technique have to rely on Gaia CRF.
%\linebreak

%   Optical positions such as those measured by the Gaia satellite are critical to establishing a 
% reference frame at comparable accuracies in different parts of the spectrum, but these missions 
% are highly episodic and not predictable; it is only VLBI, operating continuously from the 
% Earth's surface, that allows us to monitor and improve the reference frame all the time.  
% Furthermore, the Gaia mission obeys a pre-set scanning law; only VLBI can be pointed at 
% a target and observe it as needed over a wide variety of temporal samplings.  
% With Gaia running out of consumables early in the 2020s, and no other, similar missions currently 
% under development, it is only VLBI that will be able to maintain and improve the reference frame 
% over the next few decades.  VLBI, and the radio, will thus continue to be the {\it fundamental} 
% reference frame to which all other reference frames are linked, for the foreseeable future.
% %\linebreak
% 

{\bf Wide bandwidth capabilities proposed for the much more sensitive ngVLA, which includes baselines 
as long as 8900 km, will make these measurements feasible for a large number of sources, 
providing a robust statistical insight into what the causes of these offsets might be.}

% \section{Connections to the VLBI Global Observing System (VGOS)}
% 
% The International VLBI Service for Geodesy and Astrometry (IVS) is in the process of deploying an international network of small, fast slewing VLBI antennas known as the VLBI Global Observing System (VGOS).  The VGOS network is designed primarily for geodesy and thus, will be capable of observing only the brightest $\sim$10\% of quasars in the ICRF catalog. Consequently, the VGOS network will not be capable of maintaining the ICRF, including addressing issues of zonal errors between their relatively sparse sources. Minimizing these errors and those associated with intrinsic source structure and changes in the ICRF sources observed by the VGOS network will have tremendous benefit for the  network's ability to maintain its geodetic precision.
% %\linebreak
% 
% {\bf Continuous monitoring and identification of source structural changes with the %long baseline capability of the 
% ngVLA is paramount to the success of the VGOS system, which is yet another symbiotic relationship with the next generation ICRF.}
% 

\subsection{Applications of radio astrometry}

The International VLBI Service for Geodesy and Astrometry (IVS) is in the process of deploying 
an international network of small, fast slewing VLBI antennas known as the VLBI Global Observing 
System (VGOS). The VGOS network is designed primarily for geodesy. Since sensitivity of
VGOS antennas (SEFD: 2500Jy) is one order of magnitude less than that of the VLBA, it cannot
contribute to the ICRF maintenance. Therefore, it relies on imaging capabilities
of a larger array such as the ngVLA on accounting for variable source structure.
%\linebreak
 
  Another important application is space navigation. An interplanetary spacecraft
can only be observed with radio frequencies and its differential position with respect to 
a background AGN can be measured with the accuracy that is limited only with 
the absolute position of an AGN. Note that 0.1~mas error in an AGN position translates
to a position error 50~m of a spacecraft on a Martian orbit.
%\linebreak

{\bf Continuous monitoring and identification of source structural changes with the %long baseline capability of the 
ngVLA is paramount to the success of the VGOS system and space navigation, which is yet another 
symbiotic relationship with the next generation ICRF.}

\subsection{Extending the CRF to the South}
\begin{figure}[ht!]
    \centering
  \includegraphics[scale=0.6]{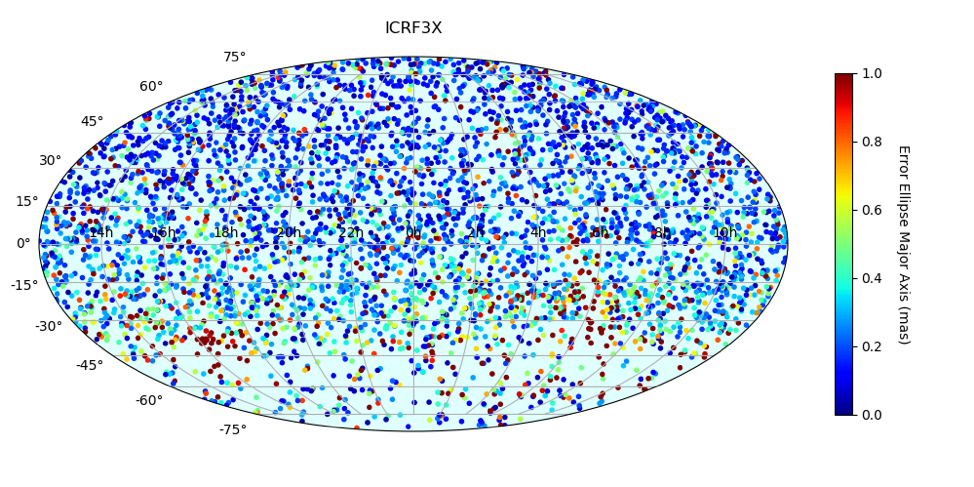}
  \caption{All sky projection of the ICRF-3 sources.}
%   \includegraphics[width=0.61\textwidth]{rfc_map.pdf}
  %  \caption{All sky projection of the RFC sources.}
    \label{fig:icrf3}
\end{figure}

 The ICRF has historically suffered from a paucity of telescopes in the southern hemisphere 
(see Figure \ref{fig:icrf3}).  The figure highlights the importance of the VLBA, since the
ICRF at declinations $>-40^\circ$ is almost entirely based on VLBA. Sources in the 
southern hemisphere observed with the VLBA have declination formal errors more than double that 
in right ascension.  North-south baselines are required to both densify and improve the accuracy of southern 
hemisphere sources.  For high frequency observations, it may be possible to join the capabilities 
of current southern hemisphere interferometers such as the Atacama Large Millimeter Array (ALMA) 
and the Large Millimeter Telescope (LMT) to the proposed ngVLA.  Some fraction of the antennas 
of the ngVLA should be geographically located as far south as possible, to 
include the southern hemisphere. This will address the source density issue, and allow us to begin 
addressing accuracy issues that are correlated with declination.  One immediate effect this 
improvement will have will be to allow us to better measure Galactic aberration using VLBI. 
%\linebreak 

{\bf A current limitation on the ICRF is the asymmetry between northern and southern hemispheres in numbers and accuracies of sources that is due to the preponderance of radio telescopes in the northern hemisphere.  Southern telescopes that could be either incorporated part of the time, or (better yet), are full-time components of a next generation VLBI system, would allow us to develop a truly isotropic reference frame, both in terms of number and accuracy of sources.}

%This would allow mitigation of this formal error discrepancy and reduce systematic and zonal errors in these regions of the sky. Finally, by increasing the source density in the southern hemisphere, we could obtain a more accurate measurement of the aberration due to the Sun's motion about the Milky Way than is currently possible with VLBI.
%%\linebreak

\section{Science with Radio Astrometry}
\subsection{Active Galactic Nuclei $-$ Studying the Dynamics of relativistic outflows}

Almost all ICRF sources are AGNs with relativistic outflows that, depending on source properties, viewing angle, and observing frequency/resolution, can be resolved. This allows for interesting synergies between the astrophysical study of AGNs, while maintaining and improving the celestial reference frame.
%\linebreak

Differences between the radio and optical AGN positions lead to one of the most spectacular results from Gaia, which is the discovery of optical relativistic jets at milli-arcsecond level \citep[1-100 mas scales;][]{pet17a}. Position offsets between VLBI (predominantly from VLBA) and Gaia observations enabled this discovery. While Gaia observations themselves have provided incremental improvements, the combination of observations of parsec-scale jets from radio observations with Gaia optical positions and photometry creates a synthetic instrument with milli-arcsecond resolution in both the radio and optical wavelengths that is able to study relativistic jets at a level that neither is possible to observe with Very Long Baseline Interferometry, nor with Gaia alone \citep{pet19}. The combined measurements enable: determination the origin of outbursts as either accretion disk, jet base, or a hot spot; determination of the ratio in the optical of synchrotron jet emission and the accretion disk; determination of the characteristic size of the optical jet; and determination of the distance of the accretion disk from the base of the jet.  It also allows us to split the populations of AGNs into sub-populations of accretion disk dominated vs. synchrotron jet dominated objects \citep{pla19a}. Within the next few years it will be possible, with existing facilities --- primarily the VLBA --- to expand the number of VLBI images of AGNs and increase accuracy of astrometric positions of targeted AGNs for combination with Gaia results.  With the advent of 25--40 m-class optical and infrared telescopes over the next decade, we expect this combination of very high resolution optical and radio data to prove extremely fruitful in observing and better understanding jets.
%\linebreak

%% I commented out this part since it does not seem relevant to astrometry. (L. Petrov)
%% Since the early days of VLBI, relativistic jets of AGNs were a primary target in order to resolve the early discovery of apparent superluminal motions \citep{sch81}. This led to the long-term monitoring of a growing sample of relativistic outflows, some of which pre-date the VLBA. In combination with the VLBA 2cm survey \citep{kel98} and Monitoring Of Jets in Active galactic nuclei with VLBA Experiments \citep[MOJAVE; e.g.,][]{lis05} a sample of the most prominent and most powerful AGNs have been monitored at mas scales for almost four decades. This growing database allowed for statistical kinematic studies of the evolution of relativistic jets, providing valuable parameters to guide relativistic magneto-hydrodynamic simulations of plasma outflows. Continuing to build this long-term database of both flaring episodes of relativistic jets, as well as quiescent periods, will allow for the study of jet precession or instabilities of the respective parsec scale emission previously not possible. This will further constrain physical conditions of such outflows \citep[e.g.,][]{sch11, lis19}. 
%%\linebreak 

The core-shift variability during AGN flares recently discovered by \citet{pla19b} %Plavin et al.(2019b) 
is one of the causes of the AGN position jitter. From the other hand, it opens the 
opportunity to study AGNs at scales finer than the size of an optically thick core.
Monitoring of core-shift changes requires a dense VLBI network that is capable
to observe at a wide range of frequencies --- and the ngVLA is the ideal instrument for that.
%\linebreak

One of the least tapped resources in astrometry and the study of relativistic outflows is polarized radio emission. A subclass of AGNs, namely BL Lac objects, were revealed to have stable polarization angle properties that are aligned with the jet axis \citep{hod18}. In this case polarimetry serves as a probe into the inner structure of relativistic outflows and can provide an accurate anchor point that combined with the continuum emission could be used as an astrometric reference point, rejecting most of the optically thin radio emission. One goal for the future ICRF is to provide position differences in orthogonal polarizations, which comes ``for free'' in terms of observing time.
%\linebreak

{\bf Enhancements to our ability to observe at high resolution in the radio---as well as access polarization information---will result in a significantly better understanding of the morphology and evolution of jets and radio-loud AGNs and better classification and modeling of source structure.  This will, in turn, further enhance our understanding of how these effects affect position accuracy for purposes of phase referencing and forming and maintaining the reference frame.}

\subsection{Parallaxes and Proper Motions}

%% I think we can omit this introduction paragraph without loss (L. Petrov)
% 
% Fundamental to our understanding of astrophysical phenomena is the ability to reliably determine distances. This is difficult within the Galaxy, since redshift can't be used and extinction is high in the plane for optical telescopes.  Reliable distance and proper motion measurements are critical to a wide research area that include the study of Galactic formation, structures, star formation, late stage stellar evolution, pulsars, star clusters, the center of our galaxy, and fundamental physics such as measurements of the Hubble constant or limits on the constancy of the gravitational constant \citep[see e.g.,][and references therein]{rei14}.
%%\linebreak

In the past decade we have seen significant progress in Galactic astronomy thanks to astrometric observations.  VLBI has provided an invaluable tool to determining model-independent distances and proper motions for pulsars. This area can particularly benefit from improvements in sensitivity due to increased recording data rates and bandwidths, as well as increased collecting area that SKA or ngVLA might provide. The measurement of pulsar proper motions is a timely topic. They provide the most stringent information on neutron star kick velocities, which affects the fraction of binary neutron star systems that remain bound and their distribution within the host galaxy: both are crucial inputs for predicting binary coalescence rates for gravitational wave experiments and the brightness of their afterglows \citep{ber14, vig18}.  
%\linebreak

Another area of great promise is using the accurate determination of absolute proper motion of regions of the Galaxy that Gaia has problems accessing due to extinction.  This includes tracing out the structure of the far side of the galaxy, and probing the Galactic center.  In addition, the next generation of VLBI will be able to make high-accuracy measurements of local group objects that will allow us to create high accuracy 3D models, providing us with a much better understanding of dynamics as well as the evolution and fate of these objects.  %of stars --??? ** Frank, can you add which objects you're referring to here? both within the center of our galaxy, as well as the proper motion of nearby galaxies in our local group that would enable a 3D model of the dynamics and evolution of our closest neighboring galaxies. %Frank, do you have a reference for this sentence here?  I'll look for one, too.
%\linebreak

Perhaps one of the most exciting radio-astronomy-related experiments over the next decade is the Event Horizon Telescope (EHT), combining mm-wavelength VLBI to observe the shadow of the event horizon of the black hole in our Galactic center. In doing so, the EHT demonstrated that astrometry at the $\rm \mu$as scale is possible. This will enable measurements of black hole positions due to perturbations from surrounding stars and/or a black hole companion \citep[e.g.,][]{lu18}. 
%\linebreak

It should be noted that the only technique currently capable of determining parallactic distances with the accuracy---and thus, the only technique capable of validating Gaia measurements---is VLBI.  For this reason alone, VLBI must be maintained and improved as an independent check on distance results from Gaia.  
%\linebreak

{\bf An expansion of capabilities at mm-wavelengths both at the VLBA and with ngVLA, will allow unprecedented capabilities to probe the innermost structures surrounding supermassive black holes as well as regions such as a Galactic center and the far side of the Galaxy that are difficult in optical frequencies due to extinction.}

\subsection{Accurate positions for Multi-wavelength science}% Frank, I have temporarily commented this section out simply due to space (5 page limit).  If you would like to add it back in, please feel free, but, if you do, could you please address Jeremy's questions regarding this section? --> Jeremy D.: "This paragraph is a little confusing:  it suggests that VLBI will be used to make radio identifications of unlocalized high-energy sources.  Wouldn?t this be the purview of a wide-area radio survey or LSST?  I?m not sure how mas-level astrometry is needed for this science; the radio and high-energy skies are not confused.  Is the idea to use radio to de-confuse the optical counterpart?" 

In astronomy, accurate source positions are needed for association 
across wavelength in the absence of correlated variability. This has 
become particularly evident in high-energy astrophysics, where source 
localization is at the arcmin level. Thus, more than one third of the 
detected $\gamma$-ray sources cannot be associated with a 
multi-wavelength counterpart \citep[e.g.,][]{ace15}, because the optical sky is 
too crowded to make statistically significant associations. However, one can take advantage of the fact that 
more than half of the known high-energy sources are radio-loud AGNs, 
with a strong physical connection between high-energy emission at $>$MeV 
energies and radio wavelengths. Thus, it is possible to establish new 
associations by probing for parsec-scale radio emission using VLBI in 
the region around unassociated $\gamma$-ray sources 
\citep[e.g.,][]{sch17}. Subsequently, through the radio positions, 
an association with an optical counterpart can be made, e.g., with Gaia 
or Pan-STARRS, which allows the establishment of a redshift measurement. 
The localization of Fermi sources from the 4FGL catalogue of $\sim\!1/2$
sources comes from VLBI and it has five orders of magnitude(!) better
accuracy than those $\gamma$-ray emitters without VLBI association.
%This example demonstrates the power of mas-scale radio positions to 
%explore one of the least known parts of the sky. It is expected that 
%the \textit{Fermi} satellite will continue operations into the next 
%decade and will be joined by the international Cherenkov Telescope Array 
%(CTA).
%\linebreak

%%In astronomy, accurate positions are needed for associating sources across wavelength. This has become particularly evident in high-energy astrophysics, where sources are localized at the arcmin level, making it particularly difficult to associate with objects at other wavelengths. Recent source catalogs published by the Fermi/Large Area Telescope collaboration \citep[e.g.,][]{ace15} list more than one third of the detected sources as unassociated, i.e. have no reliable multi-wavelength counterpart. It should be stressed that the information obtained at high-energies alone is not sufficient for astrophysical studies, since it does not provide accurate localization and no distance information.  More than half of the detected high-energy sources are radio-loud AGNs, suggesting a strong physical connection between high-energy emission at $>$MeV energies and radio wavelengths. Taking advantage of this connection, it is possible to establish new associations with dedicated radio surveys probing parsec-scale radio emission \citep[e.g.,][]{sch17}.  

{\bf Low-resolution high-energy sources can be correlated with radio sources; these can then be used, at high confidence, to identify optical and IR counterparts using Gaia and Pan-STARRS (and, in the future, LSST) observations.  This is possible because of the high-resolution of VLBI, which enables correlation between the radio and optical sources.  This correlation between high-energy sources with radio and optical counterparts allows access to significant amount of otherwise inaccessible information about the source, starting with redshift, and including many other astrophysical parameters, none of which may be available in the high-energy measurements.  This example demonstrates the power of mas-scale radio positions to explore one of the least known parts of the sky.} 

%It is expected that the Fermi satellite will continue operations into the next decade and will be joined by the international Cherenkov Telescope Array (CTA), which will rely on capabilities to perform high-resolution radio observations in order to study the multi-wavelength properties. It should be stressed that the information obtained at high-energies alone is not sufficient for astrophysical studies, since it does not provide accurate localization and no distance information.

\subsection{An Astrometric Approach to Gravitational Waves}

\begin{figure}[ht!]
\centering
%\epsscale{1.18}
\includegraphics[scale=0.6]{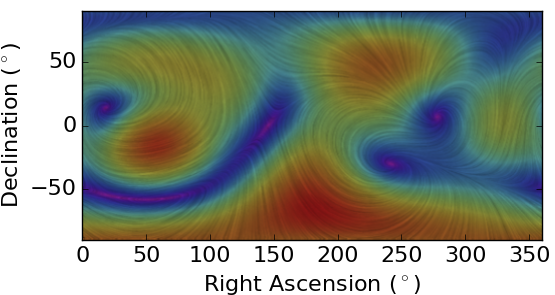}
\caption{\footnotesize%\scriptsize
  All-sky stream plot showing a random realization of apparent proper motions induced by gravitational waves with arbitrary amplitude
  (after \citealt{darling2018}).
  Streamlines indicate the vector field direction, and the colors indicate the vector amplitude, from violet (zero) to red (maximum).
  The dominant angular mode is quadrupolar.} \label{fig:streamplots}
\end{figure}

In addition to modulating length scales, gravitational waves deflect light \citep[e.g.,][]{braginsky1990,kaiser1997}.  
A dimensionless strain (fractional change in length) of $h \sim 5 \times 10^{-12}$ is
equivalent to $\sim$1 microarcsecond ($\rm \mu$as) of angular deflection.  
Stochastic gravitational waves will deflect light rays in a quadrupolar (and higher
multipole) pattern creating apparent proper motions (motions across the sky, perpendicular to the line of sight; see
Figure \ref{fig:streamplots} and \citealt{pyne1996,gwinn1997,book2011}).
Proper motions can be depicted as a vector field on the celestial sphere, which can be decomposed into
modes that resemble electric fields (curl-free E-modes) and magnetic fields (divergenceless B-modes).  Gravitational waves 
will show equal power in E- and B-modes \citep{book2011}, which can distinguish them from other signals such as
anisotropic cosmic expansion \citep{darling2014,paine2018}.
%%\linebreak

The gravitational waves that will produce extragalactic proper motions lie in the frequency range
$10^{-18}$~Hz~$< f < 10^{-8}$~Hz (H$_0$ to 0.3 yr$^{-1}$, or the observing cadence).  
This overlaps the pulsar timing frequencies \citep[$\sim$10$^{-9}$--10$^{-7}$ Hz; e.g.,][]{arzoumanian2016}
and CMB polarization frequencies \citep[$\sim$10$^{-18}$--10$^{-16}$~Hz; e.g.,][]{kamionkowski1997,seljak1997}
and uniquely spans about seven orders of magnitude in frequency between the two methods \citep{darling2018}.  Astrometry
thus provides a very large frequency window for characterization of primordial gravitational waves that is complementary to
and independent of established methods.  
%\linebreak

The cosmic energy density of gravitational waves $\Omega_{\rm GW}$ can be related to the proper motion variance $\langle\rm \mu^2\rangle$
as
\begin{equation} 
  \Omega_{\rm GW} \sim \langle\rm \mu^2\rangle/H_0^2
\end{equation}  
where H$_0 \simeq 7\times10^{-11}$~yr$^{-1} \simeq 15$~$\rm \mu$as~yr$^{-1}$ is the Hubble constant \citep{book2011}.  
Measurement of the primordial stochastic gravitational wave background from proper motions
is thus a straightforward characterization or the variance (or power) in the quadrupolar component of an all-sky proper motion signal.  
Limits have previously been obtained by \citet{gwinn1997} and \citet{titov2011} and most recently by \citet{darling2018},
$\Omega_{\rm GW} < 6.4\times10^{-3}$ (95\% confidence limit).  If {\it Gaia} can reach its expected end-of-mission performance, including
removal of current large-scale systematics, then a new limit of $\Omega_{\rm GW} < 6\times10^{-4}$ may be possible \citep{darling2018}.
%\linebreak 

{\bf
VLBI continues to offer the best outlook for a primordial gravitational wave background \citep{bower2015,darling2018b}.  An astrometry program that monitors 10,000 objects over 10 years using current VLBA-level astrometry
\boldmath{($\pm10$~$\rm \mu$as~yr$^{-1}$ per object) should reach $\Omega_{\rm GW} \sim 10^{-5}$ and will detect global
correlated signals at the level of $\sim$0.1~$\rm \mu$as~yr$^{-1}$, which is $\sim$0.7\% of $H_0$}.  Such a monitoring program would require
approximately 5 times the collecting area of the current VLBA (which monitors approximately 2,000 ICRF objects), well within the capabilities of future VLBI concepts.  }

\pagebreak
%\textbf{References}
J.~D. acknowledges support from the NSF grant AST-1411605 and the NASA grant 14-ATP14-0086.
Part of this research was carried out at the Jet Propulsion Laboratory, California Institute of Technology, under a contract with the National Aeronautics and Space Administration.
% For non-BibTex:

\end{document}